# Deep Behaviour Cloning of Model Predictive Control for Real-Time Operation of a Hydrogen–Diesel Dual-Fuel Engine

Alexander Winkler[a,*], Neeraj Naduvath Mana[a], David Gordon[b], Jakob Andert[a]

[a] *Chair of Mechatronics in Mobile Propulsion, RWTH Aachen University, Forckenbeckstr. 4, 52074 Aachen, Germany*
[b] *Department of Mechanical Engineering, University of Alberta, 116 St & 85 Ave, Edmonton, AB T6G 2R3, Canada*

## Abstract

Hydrogen-diesel dual-fuel (H2DF) combustion engines offer a promising pathway for decarbonising hard-to-electrify transport sectors, yet their highly nonlinear dynamics and coupled process variables demand constraint-aware control strategies. Model Predictive Control (MPC) meets these requirements but requires an online optimisation in every combustion cycle, which limits deployment on low-cost embedded hardware. This paper trains a feedforward deep neural network (DNN) by behaviour cloning (BC) to imitate an MPC expert, using 86,000 engine cycles of demonstration data collected at 1500 $\text{min}^{-1}$ on a modified Cummins QSB 4.5-litre hydrogen dual-fuel engine. Two variants, one with process feedback and one without, are validated experimentally. Both track unseen fast-transient load steps (3–8 bar indicated mean effective pressure, IMEP) with normalised root mean square error (NRMSE) values of 8.97% and 10.39% against the expert's 9.21%, while keeping mean NOx and particulate matter emissions at or below those of the expert. Inference takes ≤2 ms on a Raspberry Pi 400 (ARM Cortex-A72 at 2.2 GHz), including 1 ms communication latency, compared to up to 7 ms for the MPC expert, a 3.5× speedup. Open-loop profiling on a low-cost ESP32 microcontroller at 180 MHz gives 4.3 ms per inference, 4× faster than required for the 18 ms cycle window. Beyond the training range the cloned policy saturates its controls but exceeds the pressure-rise-rate limit. To the authors' knowledge, this is the first experimental BC controller for cycle-to-cycle combustion control of an internal combustion engine (ICE), trained from demonstrations recorded on the engine itself.

## 1. Introduction

Hydrogen-diesel dual-fuel (H2DF) combustion can reduce carbon-based fuel consumption in applications where full electrification remains difficult. As a short- to mid-term decarbonisation pathway, H2DF offers the prospect of cost-effective retrofitting of existing high-value diesel engine assets and fuel flexibility when hydrogen availability is limited. Compared with hydrogen fuel cells, H2DF engines can use existing compression-ignition engine platforms and tolerate less demanding hydrogen purity requirements [1, 2]. High hydrogen energy shares nevertheless introduce strongly nonlinear cycle-to-cycle dynamics, a coupled NOx–particulate matter trade-off, and safety-relevant limits on pressure rise and combustion stability [3, 4].

Model predictive control (MPC) is attractive for H2DF operation because it can handle multivariable dynamics, actuator bounds, output constraints, and rate limits within one optimization problem [5]. Its practical drawback is that a nonlinear, data-driven MPC still requires online optimization at every combustion cycle. Even when the controller is fast enough for research hardware, the repeated solution of the optimal control problem remains a barrier for deployment on low-cost embedded controllers.

A common remedy is to move the controller computation offline. Explicit MPC precomputes a control law for constrained linear-quadratic problems [6]. Neural approximations instead learn a compact representation of the MPC policy. Karg and Lucia showed that deep networks with rectified linear units can represent the piecewise-affine explicit MPC law exactly and approximate it with much smaller memory requirements [7], and demonstrated a low-cost microcontroller implementation for mixed-integer MPC [8]. Related supervised approximations have been reported for power-electronic converters [9] and for building energy management [10], and primal-dual networks have been used to learn policies that approach MPC performance while accounting for safety [11].

A further approximation route learns the optimal control action directly from expert demonstrations. Imitation learning trains such a network on these demonstrations; the supervised form used here is behaviour cloning (BC). A deep neural network (DNN) policy is trained by supervised learning to map the expert's observations to the expert's control actions, replacing the online optimization with one forward pass of the network. BC dates back to ALVINN, which learned to steer a vehicle

*Corresponding author.
*Email addresses:* alexander.winkler@rwth-aachen.de (Alexander Winkler), Mana@mmp.rwth-aachen.de (Neeraj Naduvath Mana), dgordon@ualberta.ca (David Gordon), andert@mmp.rwth-aachen.de (Jakob Andert)

from recorded driving demonstrations [12]; BC and broader imitation-learning methods have since been studied extensively in autonomous driving [13, 14, 15] and robotics [16, 17, 18]. These studies also expose the central limitation of BC. Because the cloned policy shapes the states it subsequently visits, the closed-loop state distribution drifts away from the demonstration distribution, a covariate shift under which errors compound in regions that the expert rarely visited [19]. Corrective schemes enrich the demonstration set with states that the learner is likely to visit [19, 17]. With an MPC expert, such online imitation generalised better than batch cloning in agile off-road driving [20], and data augmentation along the tube of a robust MPC reduced the number of demonstrations needed for a robust policy [21]. Both routes require access to the expert beyond the recorded demonstrations, so experimental validation of the cloned policy remains essential.

Learned approximations of predictive controllers have also been investigated in engine control. Moriyasu et al. [22] approximated a nonlinear MPC for the diesel air path with a neural network, proposed a method to limit the closed-loop degradation caused by the approximation error, and validated the resulting controller together with an unscented Kalman filter numerically and experimentally on a production engine control unit; an earlier version of the method was verified in simulation [23]. Neural controllers have likewise been applied to diesel air handling [24]. On compression-ignition engines, deep networks have also been embedded inside the predictive controller [25]. Norouzi et al. [26] did both on an engine of the same family: they augmented the optimization problem with a recurrent dynamics model and then cloned the resulting controller by deep learning, reporting a two orders of magnitude lower computation time in an experimentally validated simulation. For hydrogen/diesel dual-fuel engines, a deep-learning-based predictive injection control framework [4] and a machine-learning MPC for combustion balancing [27] have been reported. Earlier work of the authors' group by Norouzi et al. [28] integrated machine learning with MPC for imitative optimal control of a compression-ignition engine. That study established the approach in simulation using an experimentally validated diesel engine model, but did not demonstrate the imitative controller on the physical engine.

Learning-based control has also been taken to hydrogen dual-fuel operation through reinforcement learning: safe reinforcement learning has been demonstrated on an H2DF engine [29], and a hybrid scheme lets reinforcement learning adapt the load reference of a machine-learning MPC under model-plant mismatch [30]. On a single-cylinder engine in homogeneous charge compression ignition mode, a nearest-neighbour safety monitor has enabled reinforcement learning directly on the test bench [31]. These approaches learn from their own interaction with the engine; BC instead reuses expert demonstrations already generated by a validated controller, which avoids exploratory operation of the engine but inherits the expert's coverage.

The direct precursor of this work is the H2DF study by Winkler et al. [3], which developed the constraint-aware nonlinear MPC (NMPC) with a gated recurrent unit (GRU) dynamics model used here as the MPC expert, demonstrated it experimentally on the same engine platform, and provides the demonstrations cloned in this paper. The expert tracks the requested indicated mean effective pressure (IMEP), denoted by $p_{\mathrm{mi}}$, while penalising emissions, pressure-rise rate, fuel use, and control movement. In the present paper, this MPC is not re-derived.

Unlike the air-path and injection set-point approximations above, the policy studied here acts on the in-cylinder combustion process itself, cycle by cycle, and unlike the interaction-based schemes it is obtained by supervised cloning of a recurrent NMPC from nothing but the state–action pairs of demonstrations recorded on the engine. What has not been shown experimentally is whether a policy cloned from a recurrent, constraint-aware NMPC can run the combustion process of an H2DF engine within the cycle time, how far its accuracy and constraint behaviour deviate from the expert's when both are measured on the same trajectory, and how it behaves when the reference leaves the demonstrated range. The present work addresses this experimental gap through engine tests of policies with and without process feedback and an embedded runtime assessment.

The main contributions of this work are:

1. Experimental validation of a BC controller for a real H2DF internal combustion engine (ICE), trained from 86,000 cycles of demonstration data recorded on the engine with the MPC expert.

2. Direct comparison of two BC variants, with and without process feedback, against the expert H2DF MPC on the same unseen validation trajectory.

3. Embedded runtime assessment of the cloned policy on a Raspberry Pi 400 in closed loop and on an ESP32 microcontroller in open-loop profiling, together with an extrapolation test that exposes the limits of implicitly learned constraints.

The graphical abstract in Fig. 1 summarises the workflow: the H2DF MPC expert first generates demonstrations, a DNN policy is then trained by BC, and the cloned policy is finally validated on the engine. Section 2 describes the control problem and MPC data generation, Section 3 presents the DNN-based BC controller, Section 4 evaluates the cloned controller experimentally and computationally, and Section 5 concludes the paper.

## 2. Experimental Data Generation with MPC Expert

### 2.1. Experimental Setup

The experimental data has been collected from the H2DF engine platform described in detail in [3]; key engine parameters are listed in Tab. A.3 in Appendix A. The setup schematic is shown in Fig. 2. The experiments use a single fixed engine speed of 1500 min$^{-1}$, with dynamically varied transient load and challenging reference trajectories. Both the MPC expert and the BC policy run on a Raspberry Pi 400 rapid control prototyping (RCP) platform that exchanges references, feedback, and controls with the engine controller (Fig. 2). Compared with [3], only the control-relevant operating conditions and variables are repeated here. The associated MPC implementation and BC expert-demonstration resources are archived on Zenodo [32, 33].

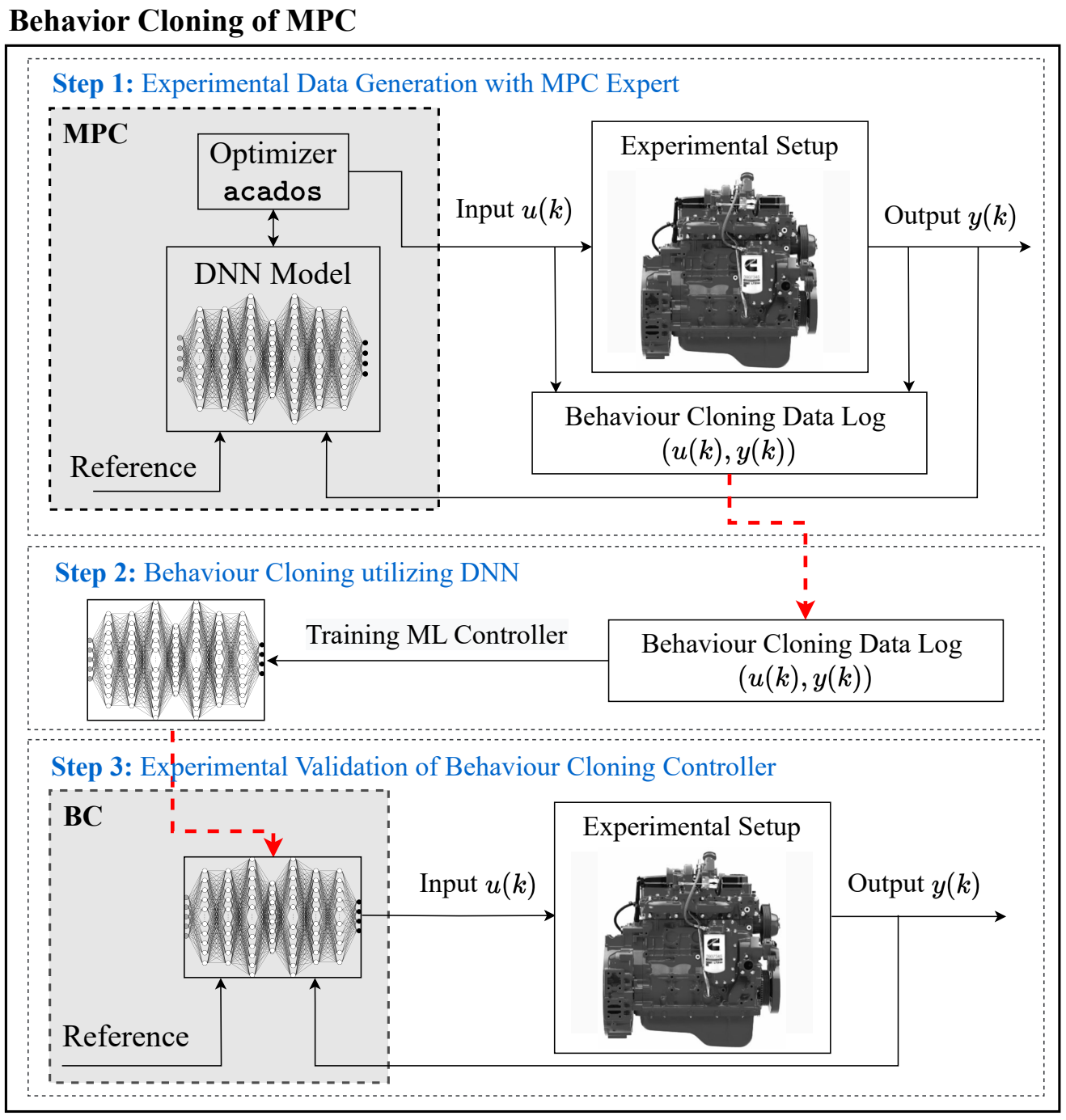

Figure 1: Modeling and controller design procedure for the BC controller based on experimental data from the MPC expert. The DNN policy is trained as a machine learning (ML) controller.

For the experiments in this paper, operating conditions outside the investigated control problem are held constant. Diesel injection is restricted to a pre-injection and a main injection. The pre-injection duration is fixed to $t_{\text{Pre}} = 0.17$ ms, corresponding to approximately 3 mg of diesel fuel; the diesel rail pressure is fixed at $p_{\text{Rail}} = 970$ bar; the hydrogen start of injection is fixed at 360 crank angle degrees (CAD) before top dead centre (bTDC), i.e. $\alpha_{\text{H2}} = 360$ CAD bTDC.

The controller sets four process inputs $u(k)$ to the engine. The index $k$ denotes the discrete combustion cycle. The controls include: duration of main injection $t_{\text{Main}}$, pre-to-main time $t_{\text{P2M}}$, start of injection (SOI) for the main diesel injection $\alpha_{\text{Main}}$ and the duration of the hydrogen injection $t_{\text{H2}}$. $u(k)$ is therefore defined as:

$$u(k) = \left[t_{\text{Main}}(k), t_{\text{P2M}}(k), \alpha_{\text{Main}}(k), t_{\text{H2}}(k)\right]^T \tag{1}$$

The pre-to-main time $t_{\text{P2M}}$ sets the start of the diesel pre-injection relative to the main injection; the injection timing over one combustion cycle and the conversion of $t_{\text{P2M}}$ from crank angle to time are detailed in [3].

The process outputs used by the MPC and by the BC evaluation are $p_{\text{mi}}$, the nitrogen oxide concentration $c_{\text{NOx}}$, the particulate matter (PM) concentration $c_{\text{PM}}$, and the maximum pressure rise rate (MPRR) $dp_{\max}$:

$$y(k) = \left[p_{\text{mi}}(k), c_{\text{NOx}}(k), c_{\text{PM}}(k), dp_{\max}(k)\right]^T \tag{2}$$

$p_{\text{mi}}$ is the tracked variable, $c_{\text{NOx}}$ and $c_{\text{PM}}$ are the regulated emissions, and $dp_{\max}$ is the combustion-safety variable: it indicates combustion noise and strain on the engine and is constrained by the MPC to protect the engine. A further quantity reported for the experiments is the hydrogen energy share (HES), the fraction of the injected fuel energy provided by hydrogen, which reflects how much diesel is substituted; the complementary diesel energy share (DES) is shown alongside it in the result figures.

The in-cylinder pressure of cylinder 1 is measured with a Kistler 6124A piezoelectric pressure transducer, from which $p_{\text{mi}}$ and $dp_{\max}$ are computed in every cycle. $c_{\text{NOx}}$ is measured with an ECM NOxCANt module and $c_{\text{PM}}$ with a Pegasor PPS-M particle sensor, both directly after the exhaust valve of cylinder 1 [3]. The PPS-M measures particulate mass using a fixed calibration for an assumed particle size; deviations from this assumption can cause reading errors in excess of 50% [34], so $c_{\text{PM}}$ carries a large measurement uncertainty. Further details on the experimental platform, measurement chain, and real-time implementation are provided in [3, 35].

### 2.2. *Expert: Model Predictive Controller*

The expert to be imitated is the H2DF MPC from [3]. It uses a learned recurrent neural-network dynamics model in the rapid MPC framework introduced in [35] and follows the data-driven MPC concept of [26]. The controller tracks the requested $p_{\text{mi}}$ while penalising $c_{\text{NOx}}$, $c_{\text{PM}}$, and $dp_{\max}$, fuel use, and aggressive control movement. The finite-horizon cost $J$ of the trajectory-tracking optimal control problem (OCP), with prediction horizon length $N$ and horizon index $i$, is

$$\begin{aligned}
J = \sum_{i=0}^{N} \Big[ & \underbrace{\|p_{\text{mi},i} - p_{\text{mi,ref},i}\|^2_{q_{p,\text{mi}}}}_{\text{Load tracking}} \\
+ & \underbrace{\|dp_{\max,i}\|^2_{q_{dp,\max}}}_{\text{Pressure-rise moderation}} \\
+ & \underbrace{\|c_{\text{NOx},i}\|^2_{q_{c,\text{NOx}}} + \|c_{\text{PM},i}\|^2_{q_{c,\text{PM}}}}_{\text{Emissions minimization}} \\
+ & \underbrace{\|t_{\text{Main},i}\|^2_{q_{t,\text{Main}}} + \|t_{\text{H2},i}\|^2_{q_{t,\text{H2}}}}_{\text{Fuel consumption and H}_2\text{ share}} \\
+ & \underbrace{\|s_i\|^1_{q_s}}_{\text{Constraint violation}} \Big] \\
+ & \sum_{i=0}^{N-1} \underbrace{\|\Delta u_i\|^2_{r_{\dot{u}}}}_{\text{Input-rate moderation}} .
\end{aligned} \tag{3}$$

Here, $\Delta u_i$ is the change of the engine-control input between consecutive horizon steps. The OCP is subject to bounds on controls, outputs, and their rates of change, handled as soft constraints with the slack vector $s_i$. The output bounds are $p_{\text{mi}} \leq 9$ bar, $c_{\text{NOx}} \leq 1200$ ppm, $c_{\text{PM}} \leq 1.5$ mg/m$^3$ and $dp_{\max} \leq 15$ bar/CAD, and the controls are limited to $t_{\text{Main}} \in [0.17, 0.50]$ ms, $t_{\text{P2M}} \in [430, 1400]$ µs, $\alpha_{\text{Main}} \in [-10, 10]$ CAD bTDC and $t_{\text{H2}} \in [1.5, 4.0]$ ms [3]. The weights $q_i$ and $r_i$ are the diagonal elements of the output and input-rate weighting matrices $Q$ and $R$, respectively; $q_s$ weights the $\ell_1$ slack penalty and $r_{\tilde{u}}$ the input changes. The weights are tunable online to adjust controller aggressiveness. The nonlinear program is solved cycle-by-cycle using a sequential quadratic programming (SQP) real-time iteration (RTI) scheme [36] within `acados` [37], with prediction horizon $N = 3$ and the high-performance interior-point method (HPIPM) quadratic-programming solver [38]. The full derivation, calibration, and validation of this MPC expert are given in [3].

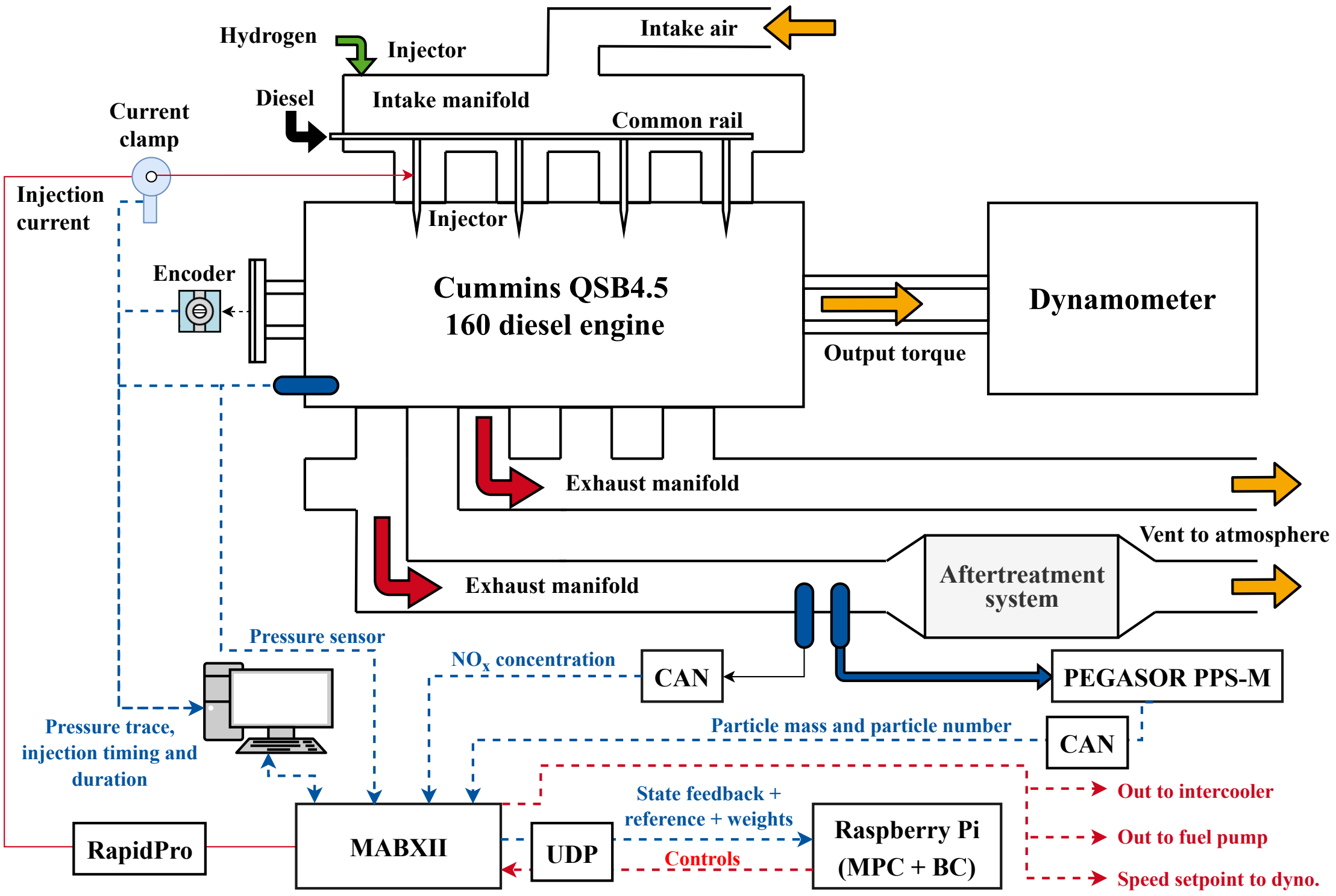

Figure 2: Experimental setup of the modified Cummins QSB 4.5L H2DF engine used for MPC data generation and BC validation, adopted from [3]. The engine is retrofitted with hydrogen port fuel injection at cylinder 1. The Raspberry Pi executes both the MPC expert during data generation and the BC policy during validation; it exchanges state feedback, references, and controls with the MicroAutoBox II (MABXII) engine controller via User Datagram Protocol (UDP). Controller area network (CAN) links transmit the emission measurements. BC: behaviour cloning.

To generate the training data, the MPC actuates the engine, thus generating a total of 86,000 cycles or data points while tracking a ramp reference trajectory on $p_{\text{mi}}$. The load trajectory randomly varied within the range of 3 to 8 bar $p_{\text{mi}}$, while keeping the rate of change of the reference limited to 0.05 bar per cycle. This ensured safe and consistent operation of the MPC, only limited by the hydrogen storage. An extract of the aforementioned randomized trajectory is shown in Fig. 3.

The load trajectory on which the BC controller was tested and evaluated is an unseen trajectory of faster dynamics, consisting of multiple quick and unconstrained jumps between static load setpoints in the range of 3 and 8 bar. The tracking results of the MPC can be seen in Fig. 4 for the process outputs.

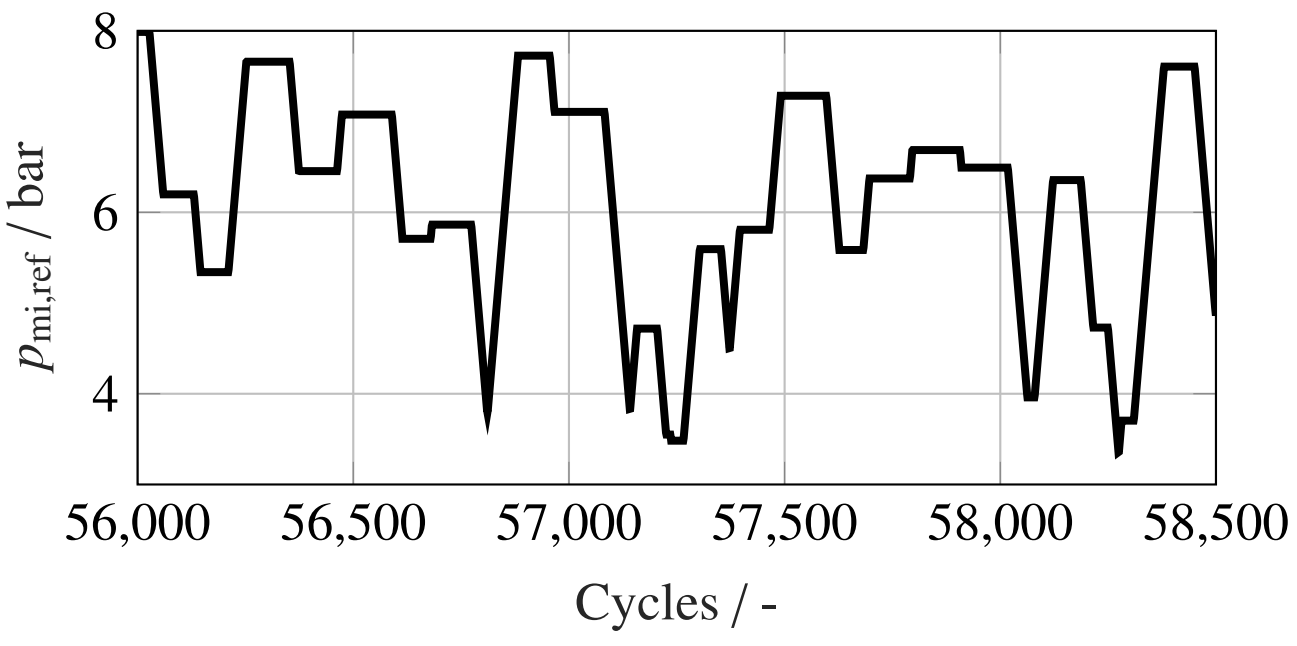

Figure 3: Extract of the $p_{\text{mi}}$ reference trajectory used for data generation with the MPC expert: randomly generated steps between 3 and 8 bar, smoothed to respect the rate limit.

The MPC struggles to track the reference at higher loads while producing minor oscillations at lower loads. Oscillations at low loads originate from incomplete opening of the hydrogen injector near its minimum activation threshold; the feedback-driven MPC amplifies this by attempting to compensate the inconsistent actuator response. A static offset at high loads reflects model-plant mismatch, mainly caused by accelerated wear of the hydrogen injector, which receives little lubrication from the pure hydrogen: the MPC's data-driven model was trained on data recorded about three months before the validation experiments, and no offset-mitigation measure such as a disturbance observer is included. Delayed settling after load steps is due to the conservative rate-of-change constraints imposed on the controls. Over this trajectory, the MPC keeps the mean $c_{\mathrm{NOx}}$ at 440 ppm and the mean $c_{\mathrm{PM}}$ at 0.12 mg/m$^3$, reaches a maximum $dp_{\mathrm{max}}$ of 13.33 bar/CAD at a mean of 4.10 bar/CAD, and attains a mean HES of 39.5% with peaks of 49.8%.

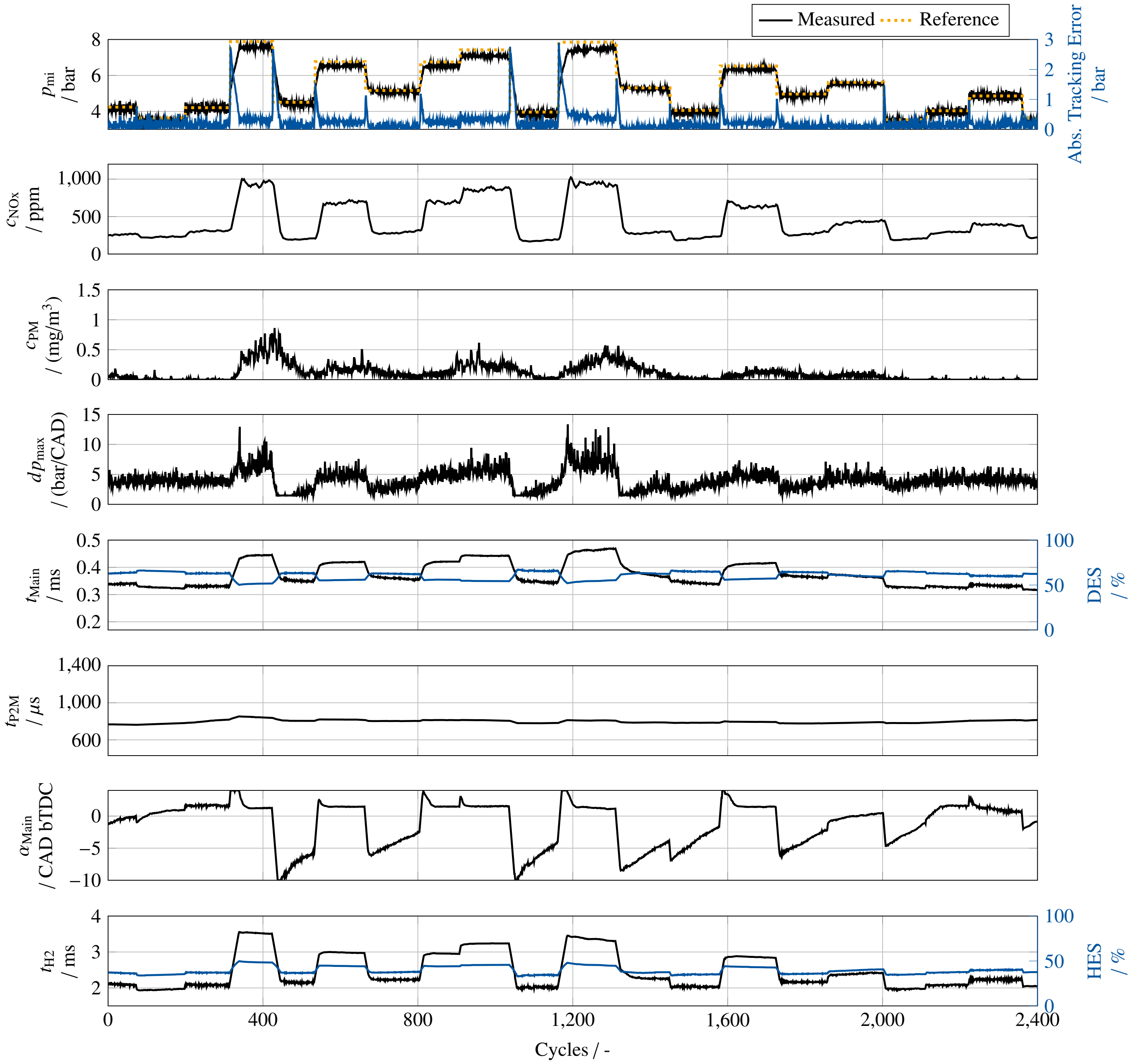

Figure 4: Process outputs and controls for the MPC expert on the dynamic reference trajectory.

## 3. Behaviour Cloning Using Deep Neural Networks

The control policy to be cloned is the MPC. The MPC control actions are used as the target outputs, and the MPC-relevant inputs are used as the DNN inputs. To reduce complexity, the input dimension is simplified to the four most important signals: the reference trajectory $p_{\mathrm{mi,ref,i}}$ for the next three cycles, $i \in \{1, 2, 3\}$, aligning with the prediction horizon length $N = 3$ of the MPC expert, and the feedback of the measured and calculated $p_{\mathrm{mi}}(k-1)$. The DNN inputs $\boldsymbol{X}_{\mathrm{DNN}}(k)$ are defined as:

$$\boldsymbol{X}_{\mathrm{DNN}}(k) = \left[p_{\mathrm{mi,ref,1}}(k), p_{\mathrm{mi,ref,2}}(k), p_{\mathrm{mi,ref,3}}(k), p_{\mathrm{mi}}(k-1)\right]^T, \quad \boldsymbol{X}_{\mathrm{DNN}}(k) \in \mathbb{R}^4. \tag{4}$$

For the later replacement of the MPC by the DNN, the exact MPC's control outputs $\boldsymbol{u}(k)$ defined in Eq. 1 are chosen to be the neural networks training outputs $\boldsymbol{U}_{\mathrm{DNN}}(k)$ to be predicted:

$$\begin{aligned}\boldsymbol{U}_{\mathrm{DNN}}(k) &= u(k),\\ &= \left[t_{\mathrm{Main}}(k), t_{\mathrm{P2M}}(k), \alpha_{\mathrm{Main}}(k), t_{\mathrm{H2}}(k)\right]^T,\\ \boldsymbol{U}_{\mathrm{DNN}}(k) &\in \mathbb{R}^4. \end{aligned} \quad (5)$$

With these signals, BC replaces the MPC policy of Section 2.2 by a DNN policy that is evaluated in one forward pass. Its state and control spaces are subsets of those of the expert,

$$\boldsymbol{X}_{\mathrm{DNN}} \subseteq \mathbb{R}^{n_x}, \qquad \boldsymbol{U}_{\mathrm{DNN}} \subseteq \mathbb{R}^{n_u}, \quad (6)$$

with $n_x = n_u = 4$ for the signals defined in Eq. 4 and Eq. 5. The cloned policy $\pi_\theta$, parameterised by the network parameters $\theta$, maps a state to a control action,

$$\pi_\theta : \boldsymbol{X}_{\mathrm{DNN}} \rightarrow \boldsymbol{U}_{\mathrm{DNN}}, \qquad \boldsymbol{U}_{\mathrm{DNN}}(k) = \pi_\theta(\boldsymbol{X}_{\mathrm{DNN}}(k)), \quad (7)$$

where $k$ is the combustion cycle index. The policy is trained on the demonstrations recorded while the MPC actuates the engine,

$$\mathcal{D} = \{(x_i, u_i)\}_{i=1}^{N_{\mathrm{data}}}, \qquad x_i \in \boldsymbol{X}_{\mathrm{DNN}}, \quad u_i \in \boldsymbol{U}_{\mathrm{DNN}}, \quad (8)$$

in which $u_i$ is the control action the expert computed for the state $x_i$. Training is the supervised problem of finding the parameters that minimise the deviation between cloned and expert actions,

$$\theta^* = \arg\min_\theta \sum_{i=1}^{N_{\mathrm{data}}} L(\pi_\theta(x_i), u_i), \quad (9)$$

with the mean squared error as loss function $L$. Unlike the expert, the cloned policy carries no explicit model, cost function, or constraints; it reproduces the expert's input-output behaviour on the demonstrated distribution.

The DNN trained on these demonstrations is a feedforward neural network (FFNN) with an input layer, multiple fully connected layers using Rectified Linear Unit (ReLU) activation functions, and an output layer; its compact architecture is shown in Fig. 5.

The DNN used in this work comprises a total of 21,460 learnable parameters during training. A learned model inside the MPC, as used in Section 2.2 and in [26], can make the prediction step cheaper to evaluate than a physically derived one, but it does not remove the overhead of the online iterative optimization, because the network is evaluated repeatedly within every solver iteration. Therefore, replacing the iterative optimization with one DNN forward pass reduces the deployed controller to a single forward inference of the network.

The dataset, consisting of a total number of 86,000 engine cycles or samples ($N_{\mathrm{data}}$) of three different measurement runs, is sequentially divided into training, validation, and test datasets for each of the measurement runs. The total dataset is thus split with proportions of 80%, 15%, and 5%, respectively. The data distribution of the input with the highest variance, the feedback $p_{\mathrm{mi}}(k-1)$, is shown in Fig. 6. The distributions are consistent across the splits, ensuring reliable comparability for training and evaluation. All data is normalized using the total dataset's minimum and maximum values.

The training hyperparameters are listed in Table B.4. The prediction results of the DNN in the time-domain on the unseen test dataset are shown in Fig. C.10 (Appendix), showing good correspondence with the MPC data. The average normalised root mean square error (NRMSE) across all four prediction outputs is 5.39%. The two most important controls, the injection durations of diesel ($t_{\mathrm{Main}}$) and hydrogen ($t_{\mathrm{H2}}$), achieve NRMSE values of 4.48% and 3.80% respectively on the unseen test dataset, and the pre-to-main time $t_{\mathrm{P2M}}$ reaches 2.97%, while the main injection timing $\alpha_{\mathrm{Main}}$ shows a higher NRMSE of 10.29% due to the high variance of the target combined with limited discriminability in the input features.

## 4. Experimental Validation of Behaviour Cloning Controller

### 4.1. Controller Definition & Experimental Setup

To actuate the engine, the control policy $\pi_\theta$ is integrated into the real-time control model using `MATLAB/Simulink` code-generation blocks. The BC controller receives the same reference interface as the MPC expert and returns the four injection commands defined in Eq. 1.

Two variants of the cloned policy are evaluated on the same unseen trajectory. Variant A uses the full input vector of Eq. 4, including the measured load feedback $p_{\mathrm{mi}}(k-1)$. Variant B omits this feedback input (Fig. 5) and therefore acts purely feedforward on the load reference, without any process measurement. Both variants are compared against the MPC expert on a single validation run of 2400 engine cycles at 1500 $\mathrm{min}^{-1}$. Load tracking is evaluated by the mean absolute error (MAE), the root mean square error (RMSE), and the NRMSE, which is normalised by the range of the reference trajectory (4.35 bar, from 3.54 to 7.88 bar).

### 4.2. Engine Load Tracking Performance

The experimental results of Variant A are displayed in Fig. 7. The unseen reference trajectory of $p_{\mathrm{mi}}$ consists of random fast load jumps between 3 and 8 bar. These fast unconstrained steps are absent from the rate-limited training trajectory of Fig. 3 and therefore challenge the cloned policy. While the reference is followed accurately, a smooth engine control is achieved. As for the MPC, a slight static offset remains at high loads, caused by the injector-wear-related model-plant mismatch described in Section 2.2; the BC controller inherits this behaviour from its expert.

The performance of the BC controller is compared to the MPC expert (see Fig. 4); Table 1 lists the metrics of Variant A against the expert. Regarding the accuracy of the reference trajectory tracking for $p_{\mathrm{mi}}$, the BC controller achieves an NRMSE of 8.97%, slightly below the MPC's 9.21%, whereas its MAE is higher (0.311 vs. 0.242 bar). The MPC's rate-limited controls settle slowly after the fast load steps, producing fewer small deviations but larger transient errors, which weigh more heavily in the RMSE. The cloned policy carries no such rate limit: it changes $t_{Main}$ by up to 0.067 ms and $t_{H2}$ by up to 0.626 ms from one cycle to the next, against 0.008 ms and 0.120 ms for the MPC, so it reproduces the expert's steady-state control levels but not its rate-limited transitions. However, in terms of emissions, the BC controller maintains $c_{NOx}$ at an average level of 424 ppm and $c_{PM}$ at 0.06 mg/m$^3$, meeting or improving upon the MPC's emission levels of 440 ppm and 0.12 mg/m$^3$, respectively. The lower $c_{PM}$ is attributed mainly to the general measurement uncertainty of the PM sensor (Section 2.1) rather than to a systematic advantage of the BC controller. Both emission figures are mean concentrations rather than load-normalised values, so part of the difference follows from the slightly lower load delivered by the BC controller at high references. The BC controller therefore tracks this unseen fast reference while keeping mean $c_{NOx}$ and $c_{PM}$ at or below the expert's levels; its mean $dp_{max}$, however, is 5.7% higher. The mean HES of 42.26% exceeds the expert's 39.50% by 7.0%, while its maximum stays 6.0% below, so the cloned policy substitutes at least as much diesel by hydrogen as the expert.

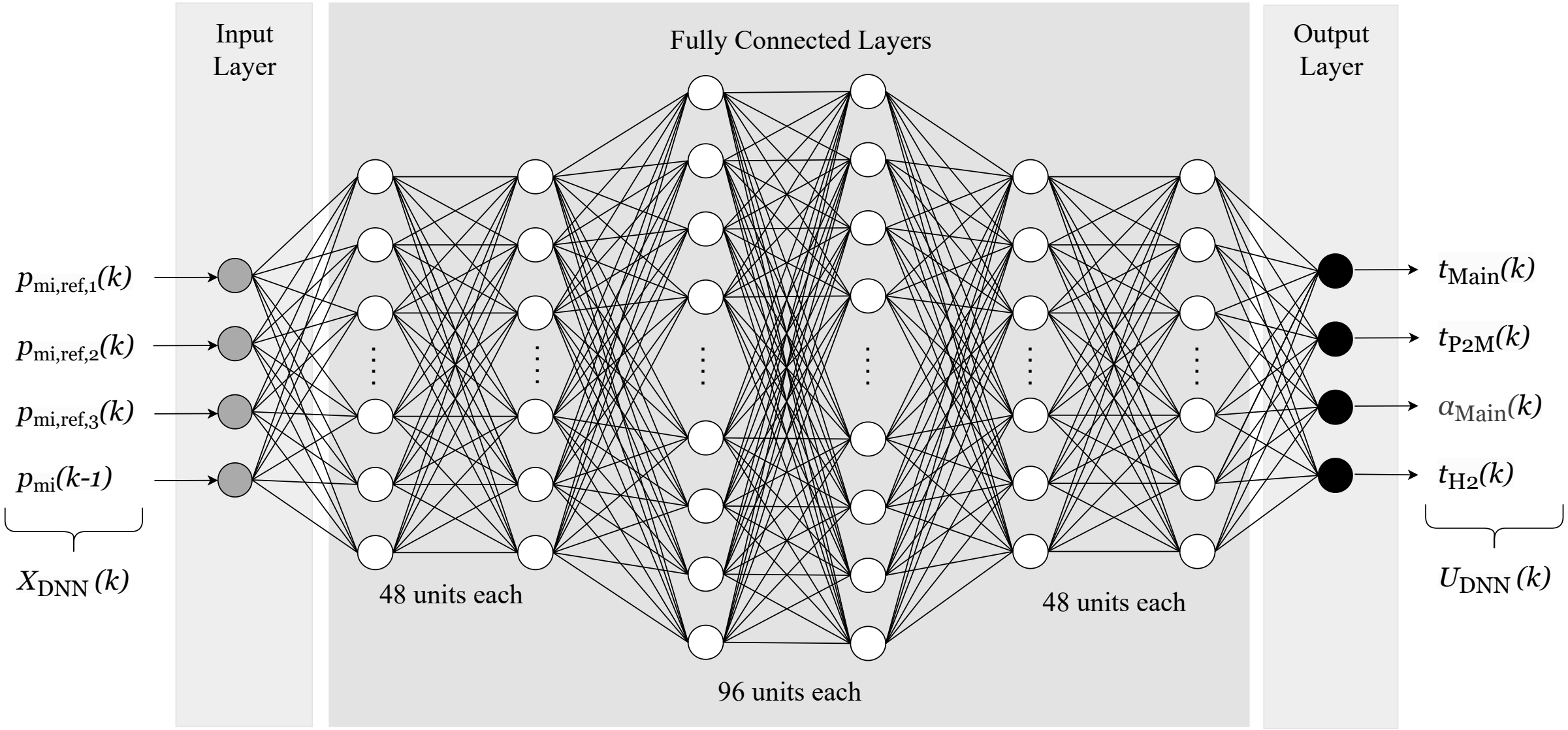

Figure 5: Fully connected DNN architecture with Rectified Linear Unit activation functions and 21,460 learnable parameters in total. Variant A includes the feedback variable $p_{mi}(k-1)$, Variant B excludes it (not shown separately).

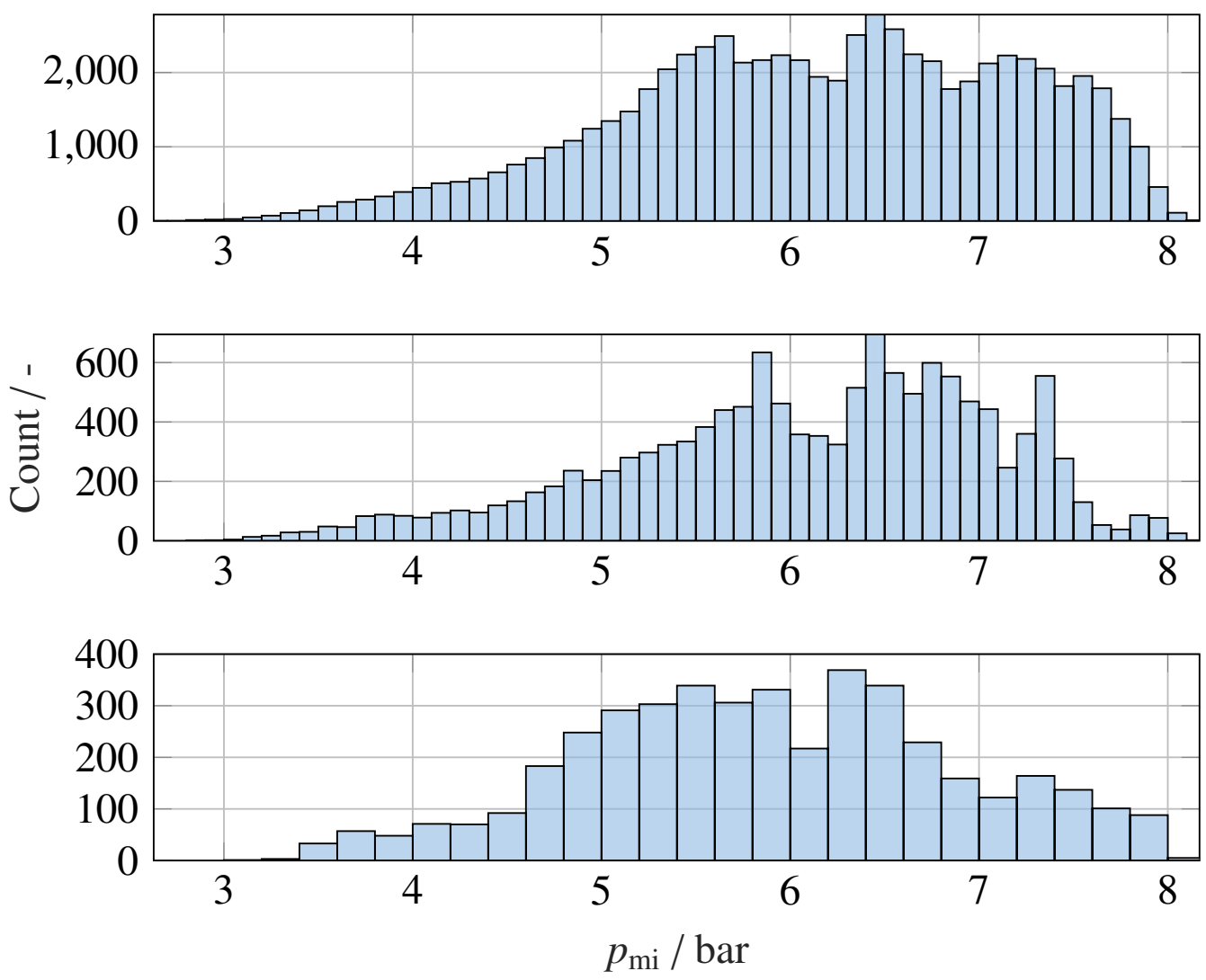

Figure 6: Data distribution of the DNN input $p_{mi}(k-1)$. Upper plot: training dataset, middle plot: validation dataset, lower plot: unseen test dataset. Total data points ($N_{data}$): 86,000.

Variant B receives the same unseen reference trajectory but no process measurement; its metrics are listed in Table 2 and the full output traces in Appendix C (Fig. C.11). Without feedback, tracking degrades modestly (NRMSE 10.39% vs. 8.97% for feedback BC and 9.21% for MPC), mainly through small low-load offsets. Mean $c_{NOx}$ and $c_{PM}$ remain below the expert's levels while the mean $dp_{max}$ is 3.5% higher and the mean HES is 6.8% higher, indicating that the reference alone carries enough information for the policy to reproduce the expert's fuel path at this operating point.

Table 1: Variant A (with feedback) versus the MPC expert on the unseen validation trajectory (2400 cycles, 1500 min$^{-1}$). Δ: relative deviation from the MPC. NRMSE is normalised by the 4.35 bar range of the reference trajectory.

| **Metric** | **Variant A** | **MPC** | **Δ vs. MPC** |
|---|---|---|---|
| **$p_{mi}$ (load tracking)** | | | |
| MAE / bar | 0.311 | 0.242 | +28.7% |
| RMSE / bar | 0.390 | 0.400 | −2.6% |
| NRMSE / % | 8.97 | 9.21 | −2.6% |
| **$c_{NOx}$ (emissions)** | | | |
| max / ppm | 927 | 1024 | −9.5% |
| mean / ppm | 424 | 440 | −3.6% |
| **$c_{PM}$ (emissions)** | | | |
| max / mg m$^{-3}$ | 0.52 | 0.86 | −40.0% |
| mean / mg m$^{-3}$ | 0.06 | 0.12 | −48.5% |
| **$dp_{max}$ (pressure rise rate)** | | | |
| max / bar CAD$^{-1}$ | 12.08 | 13.33 | −9.4% |
| mean / bar CAD$^{-1}$ | 4.33 | 4.10 | +5.7% |
| **HES** | | | |
| max / % | 46.86 | 49.83 | −6.0% |
| mean / % | 42.26 | 39.50 | +7.0% |

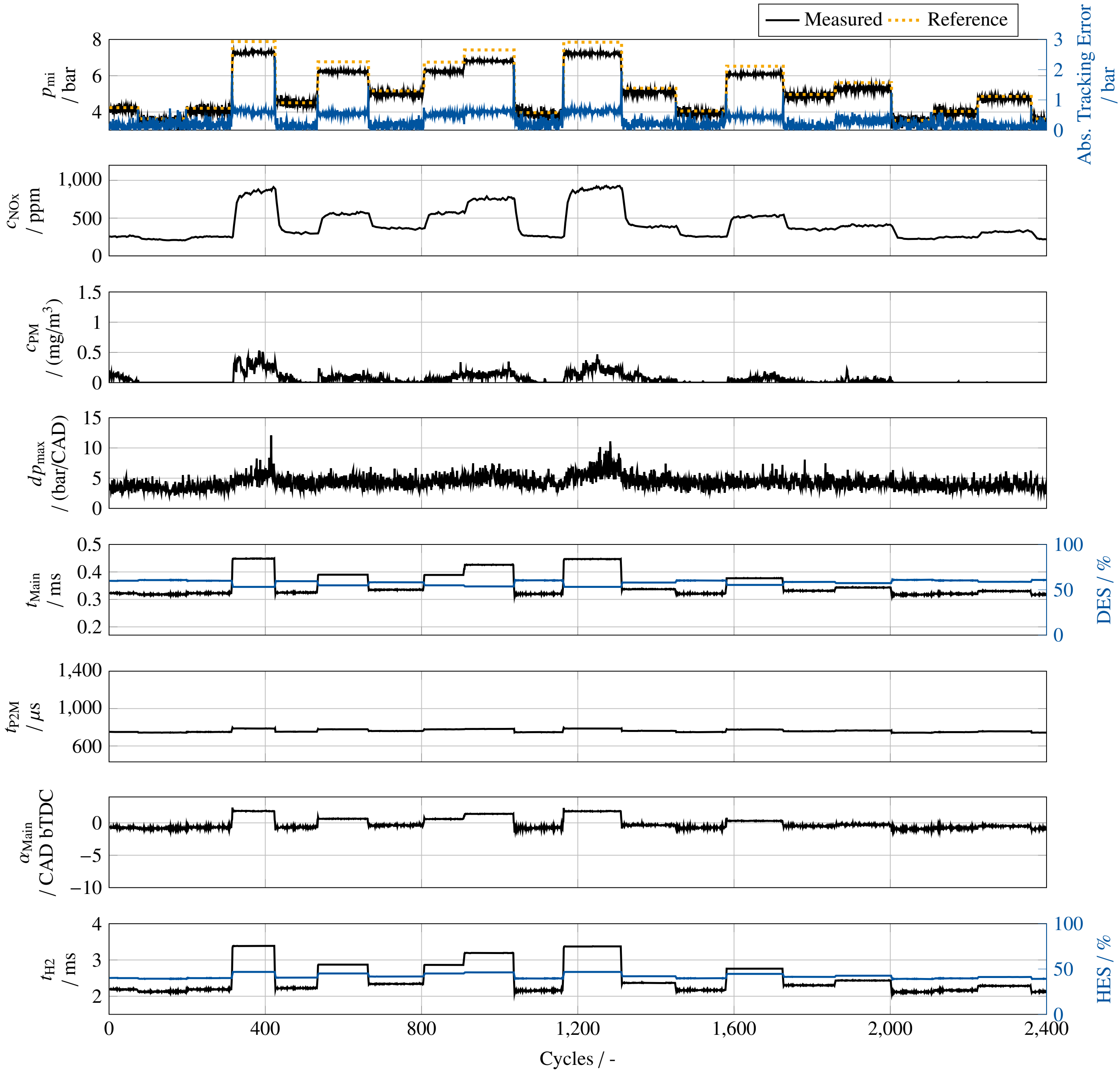

Figure 7: Process outputs and controls for the BC controller with feedback (Variant A) on an unseen dynamic reference trajectory.

Both BC variants successfully replicate the MPC expert: load-tracking NRMSE is within 1.2 percentage points of the MPC, mean $c_{NOx}$ is reduced by 3.6–7.5% and mean $c_{PM}$ by 24.7–48.5%, the latter attributed mainly to PM sensor measurement errors, and the HES stays within ±7% of the expert's, confirming that BC captures the MPC's fuel strategy. The mean $dp_{max}$ is up to 5.7% higher for BC than for the MPC.

### *4.3. Extrapolation and Constraint Behaviour*

Given the absence of formal safety guarantees in data-driven controllers, it is critical to assess how the BC policy behaves when faced with out-of-distribution inputs. Fig. 8 shows Variant A's response when the $p_{mi}$ reference is pushed to 9.5 bar, beyond the 8 bar upper bound of the training distribution. This test was performed with Variant A only, in a single run.

The controller does not violate the implicitly learned bounds on its outputs; instead, it maximises both diesel and hydrogen injection durations in an attempt to reach the requested load, and $c_{NOx}$ and $c_{PM}$ rise accordingly. This indicates stable extrapolation of the learned policy for the tested load step. Neverthe-less, the high fuel quantities lead to elevated $dp_{max}$ values, confirming that the $dp_{max}$ constraint was not reliably internalised from the training data. This outcome is expected given the limited representation of near-constraint operation in the MPC-generated demonstration set.

Table 2: Variant B (without feedback) versus the MPC expert on the unseen validation trajectory (2400 cycles, 1500 min$^{-1}$). Δ: relative deviation from the MPC. NRMSE is normalised by the 4.35 bar range of the reference trajectory.

| Metric | Variant B | MPC | Δ vs. MPC |
|---|---|---|---|
| **$p_{mi}$ (load tracking)** | | | |
| MAE / bar | 0.391 | 0.242 | +61.7% |
| RMSE / bar | 0.452 | 0.400 | +12.8% |
| NRMSE / % | 10.39 | 9.21 | +12.8% |
| **$c_{NOx}$ (emissions)** | | | |
| max / ppm | 884 | 1024 | −13.7% |
| mean / ppm | 407 | 440 | −7.5% |
| **$c_{PM}$ (emissions)** | | | |
| max / mg m$^{-3}$ | 0.59 | 0.86 | −31.7% |
| mean / mg m$^{-3}$ | 0.09 | 0.12 | −24.7% |
| **$dp_{max}$ (pressure rise rate)** | | | |
| max / bar CAD$^{-1}$ | 9.71 | 13.33 | −27.2% |
| mean / bar CAD$^{-1}$ | 4.24 | 4.10 | +3.5% |
| **HES** | | | |
| max / % | 46.78 | 49.83 | −6.1% |
| mean / % | 42.18 | 39.50 | +6.8% |

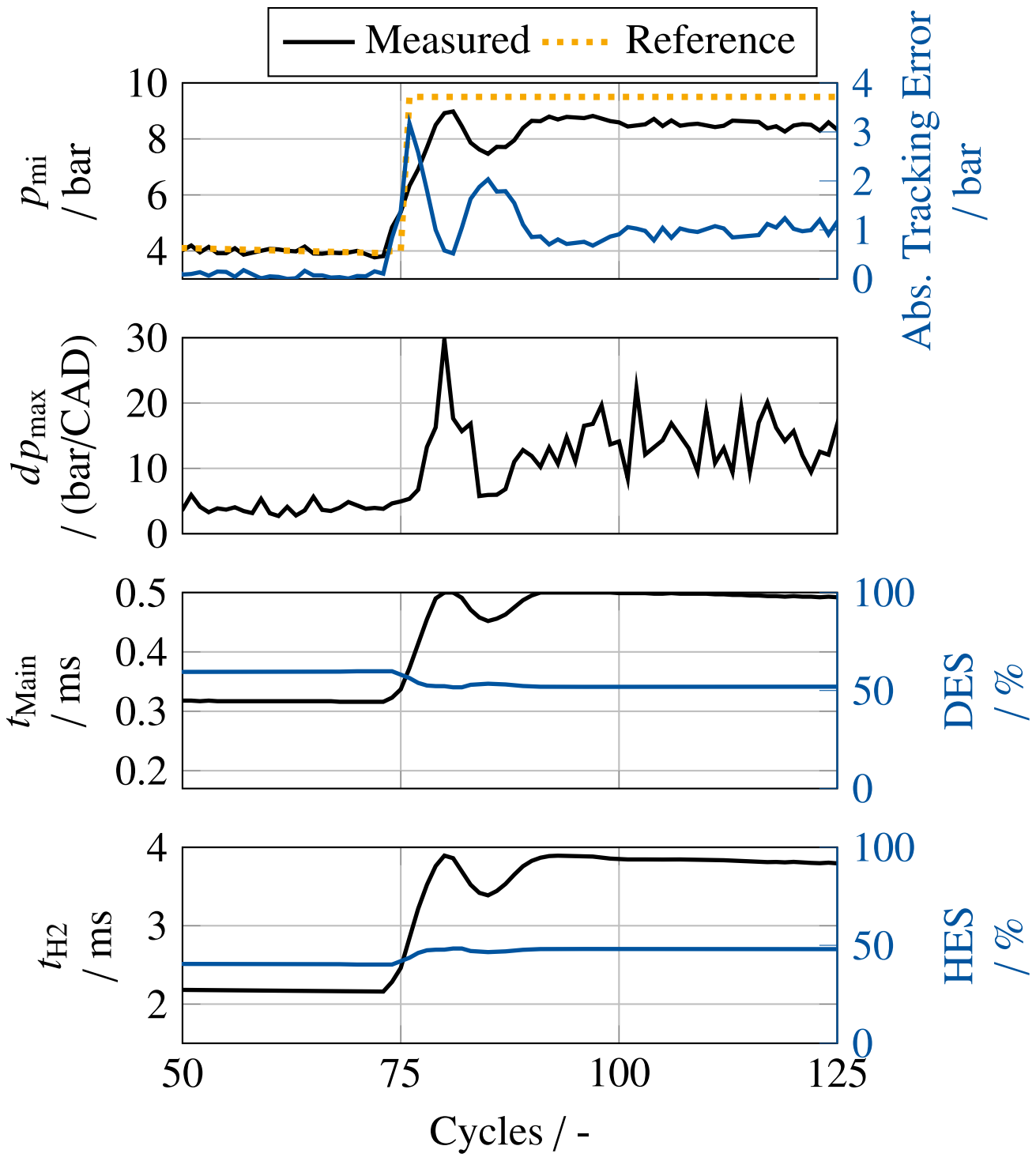

Figure 8: BC controller (Variant A) extrapolation: response when $p_{mi,ref}$ exceeds the training range (9.5 bar vs. 8 bar maximum). Controls saturate; the $dp_{max}$ constraint is violated. The remaining outputs and controls are omitted.

### 4.4. *Computational Performance*

At 1500 min$^{-1}$ one four-stroke engine cycle lasts 80 ms, but the combustion metrics of the current cycle only become available late in the expansion stroke, which leaves an 18 ms window for the controller to compute the commands for the next cycle [3]. The BC controller consistently achieved computation times of ≤2 ms on the Raspberry Pi 400 (ARM Cortex-A72 at 2.2 GHz), including the 1 ms UDP communication latency. The MPC achieved a run-time of up to 7 ms under the same conditions, yielding a 3.5× speedup for the BC controller. Both measurements include the 1 ms UDP overhead; the net DNN inference time is thus below 1 ms, consistent with the two orders of magnitude reduction in computation time reported in simulation by Norouzi et al. [26]. This results in at least a 9-fold real-time margin for the 18 ms available computation window.

For an accurate definition of the DNN-based controller's real net computational time without any communication overhead and limitation due to low base-rates, an additional profiling is conducted. The low-cost ESP32 microcontroller with 180 MHz base frequency and limited memory is used as a profiling target to assess whether the BC policy could run on resource-constrained hardware. The controller was not used to operate the engine directly because the test bench is built around the flexible RCP integration described in [35, 3], where the generated controller interfaces with the dSPACE hardware, safety logic, and measurement infrastructure.

Multiple DNNs of different size and unit type were integrated into an open-loop simplified simulative environment to assess the net calculation time. The results of the study are displayed in Fig. 9. The network types differ in their recurrent computations. A GRU uses update and reset gates with a hidden state, whereas a long short-term memory (LSTM) cell maintains a separate cell state and hidden state with additional gating; an FFNN has no recurrent state [39, 40, 41]. This simpler structure can make GRUs less expensive to evaluate than LSTMs [42], which is why a GRU was chosen for the dynamics model of the MPC expert. In the present benchmark, maximum execution time increases approximately linearly with the number of learnable parameters within each network type, with the observed ordering LSTM > GRU > FFNN for comparable parameter counts. The recurrent state updates, gates, and activation functions add inference work; the measured slopes also depend on the implementation and hardware. The BC networks of Variant A (21,460 learnable parameters, 42,580 floating-point operations (FLOPs) per inference) and Variant B (21,412) were measured at 4.30 ms and 4.50 ms maximum execution time on the ESP32. Either network therefore occupies roughly a quarter of the 18 ms computation window, 4× faster than required. The marker labelled MPC model in Fig. 9 refers to the GRU dynamics model of the expert (2.23 ms); it quantifies the forward evaluation of the prediction model only, not the SQP iteration and the interior-point solve that the MPC additionally performs every cycle. The complete MPC was not profiled on the ESP32 because its solver exceeds the memory of the microcontroller; its measured run-time on the Raspberry Pi 400 is reported above.

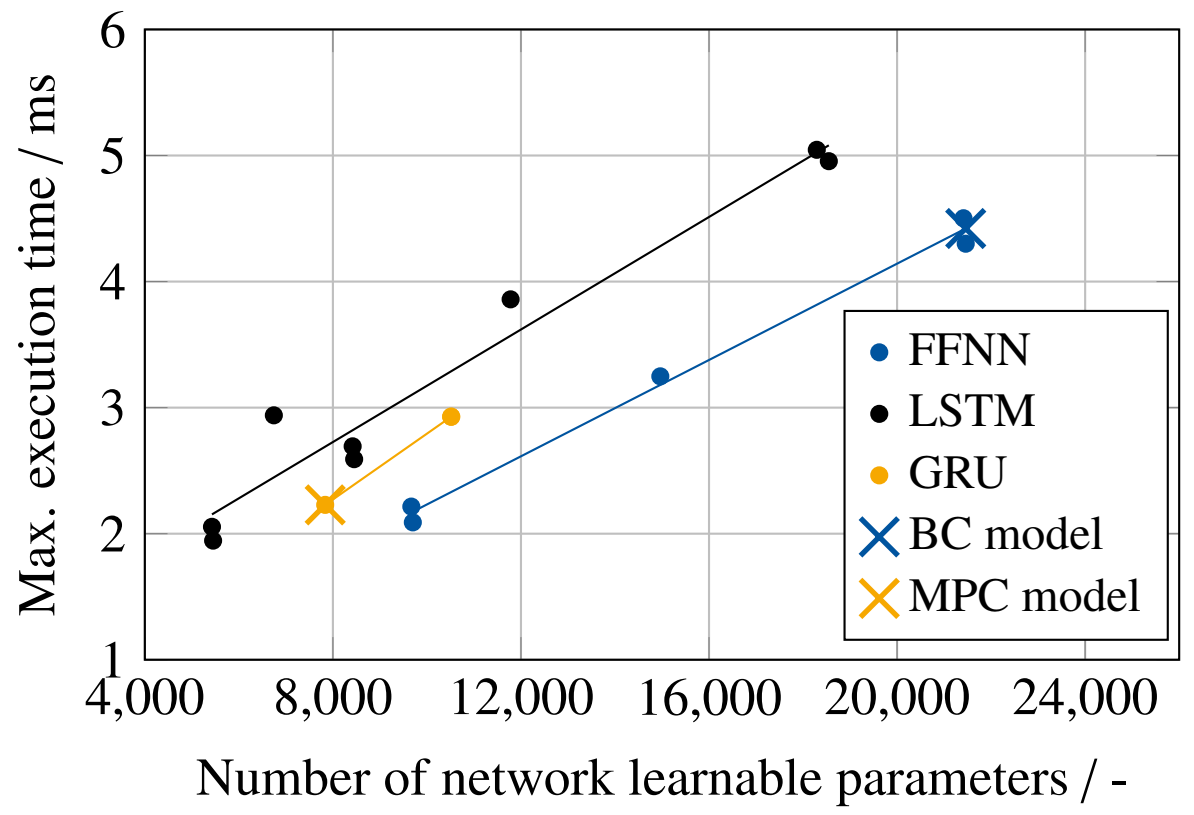

Figure 9: Profiling results on low-cost embedded controller ESP32 with 180 MHz. Dots show individual networks and solid lines the linear trend of each network type; cross markers indicate the BC DNN and the GRU dynamics model of the MPC expert, both placed on the trend of their type. The BC marker represents both BC variants, which are treated as equal. The MPC marker covers the dynamics-model evaluation only and excludes the online optimization. GRU: Gated Recurrent Unit, LSTM: Long Short-Term Memory, FFNN: Feedforward Neural Network.

## 5. Conclusions

This work presented the first experimental BC controller for cycle-to-cycle combustion control of an ICE, trained from demonstrations recorded on the engine itself. A DNN trained on 86,000 engine cycles of MPC expert demonstrations on an H2DF engine replaced the online optimisation of the MPC and acts within one combustion cycle. Two variants, with and without process feedback, were compared directly against the MPC expert on the same unseen trajectory, and the cloned policy was assessed on embedded hardware, in closed loop on a Raspberry Pi 400 and in open-loop profiling on an ESP32 microcontroller, together with an extrapolation test beyond the training range.

All tests used a single fixed engine speed of 1500 $\text{min}^{-1}$ with dynamically varied transient load $p_{\text{mi}}$. The feedback variant achieved an NRMSE of 8.97%, closely matching the MPC's 9.21%, while the feedback-free variant achieved 10.39%. Both variants kept mean $c_{\text{NOx}}$ and $c_{\text{PM}}$ at or below the expert's levels, while their mean $dp_{\text{max}}$ was 5.7% and 3.5% higher. A single extrapolation test with Variant A, in which the load reference $p_{\text{mi,ref}}$ was raised to 9.5 bar against an 8 bar training maximum, showed control saturation but a violation of the $dp_{\text{max}}$ constraint. Operating without any process measurement therefore costs 1.4 percentage points of tracking accuracy, which quantifies the performance attainable from the reference alone. These tests establish performance for the investigated fixed-speed conditions, not across an engine speed-load map.

The primary advantage of the BC controller is computational efficiency: inference times below 2 ms on a Raspberry Pi 400, including 1 ms communication latency, represent a 3.5× speedup over the MPC. Open-loop profiling on an ESP32 microcontroller gave 4.30 ms per inference for the trained network (42,580 FLOPs), 4× faster than required for the cycle window, indicating that the arithmetic cost of the policy is compatible with low-cost embedded hardware.

Despite its demonstrated performance, some limitations remain. The presented approach does not solve the well-known BC distribution-shift problem: once deployed in closed loop, the controller may visit states that were rare or absent in the expert dataset. In the single extrapolation test of Section 4.3, load tracking remained stable under such a shift, but the output constraints of the expert were exceeded. No formal constraint guarantees are provided, and the purely data-driven policy only internalises safety-relevant limits indirectly from the demonstrations. The method is also tied to the operating point and to the expert demonstrations used for training; broader speed and load coverage would require additional data. Future work should address these limitations through noise injection into the demonstrations, dataset aggregation, or constraint-augmented architectures [17, 19]. A supervisory safety filter that clips the cloned controls to the expert's constraint set would bound the behaviour outside the demonstrated range without retraining. Feeding the measured $dp_{\text{max}}$ back into the policy, so that the constraint dynamics enter the input space explicitly, is a further option that remains untested here.

## Author Contributions

**Alexander Winkler**: Conceptualization, Methodology, Software, Validation, Formal Analysis, Investigation, Data Curation, Writing – Original Draft, Visualization. **Neeraj Naduvath Mana**: Methodology, Software, Investigation, Writing – Review & Editing. **David Gordon**: Resources, Investigation, Writing – Review & Editing, Supervision, Funding Acquisition, Project Administration. **Jakob Andert**: Supervision, Writing – Review & Editing, Funding Acquisition, Project Administration.

## Funding

This work was supported by the Deutsche Forschungsgemeinschaft (DFG, German Research Foundation) as part of the Research Group FOR 2401 "Optimization-based Multiscale Control for Low Temperature Combustion Engines". Additional support was provided by the RWTH Aachen University–University of Alberta Junior Research Fellowship, the Natural Sciences and Engineering Research Council of Canada Discovery Grant RGPIN-2024-04990, and a Mitacs Globalink Research Internship. The funding sources had no involvement in the study design; data collection, analysis, or interpretation; manuscript preparation; or the decision to submit the article.

## Declaration of Competing Interest

The authors declare that they have no known competing financial interests or personal relationships that could have appeared to influence the work reported in this paper.

## Acknowledgments

The authors gratefully acknowledge the contributions of all members of Research Group FOR 2401 for stimulating discussions. The authors thank Nikolaus Schwind for his support during this work, and thank the technical staff of the Department of Mechanical Engineering at the University of Alberta for their support during engine test bench operations, especially Javad Kheyrollahi and Edward Sperling.

## Data Availability

The MPC implementation and closed-loop simulation resources associated with the expert controller are available on Zenodo [32]. The BC scripts, controller resources, and MPC expert demonstration dataset used in this paper are available on Zenodo [33].

## Declaration of generative AI and AI-assisted technologies in the manuscript preparation process

During the preparation of this work, the authors used Codex 5.5 for grammar, formatting, and language editing. After using this tool, the authors reviewed and edited the content as needed and take full responsibility for the content of the published article.

## Abbreviations

| Abbreviation | Description |
|---|---|
| BC | Behaviour cloning |
| CAD | Crank angle degree |
| DES | Diesel energy share |
| DNN | Deep neural network |
| FFNN | Feedforward neural network |
| GRU | Gated recurrent unit |
| H2DF | Hydrogen-diesel dual-fuel |
| HES | Hydrogen energy share |
| HPIPM | High-performance interior-point method |
| ICE | Internal combustion engine |
| IMEP | Indicated mean effective pressure |
| LSTM | Long short-term memory |
| MAE | Mean absolute error |
| MPC | Model predictive control |
| MPRR | Maximum pressure rise rate |
| NMPC | Nonlinear model predictive control |
| NRMSE | Normalised root mean square error |
| OCP | Optimal control problem |
| PM | Particulate matter |
| RCP | Rapid control prototyping |
| RMSE | Root mean square error |
| RTI | Real-time iteration |
| SOI | Start of injection |
| SQP | Sequential quadratic programming |
| UDP | User Datagram Protocol |

## Symbols

| Symbol | Description |
|---|---|
| *Behaviour cloning policy* | |
| $\pi_\theta$, $\theta$ | Neural-network control policy and its trainable parameters |
| $\boldsymbol{X}_{\text{DNN}}$, $\boldsymbol{U}_{\text{DNN}}$ | Policy input (state) space and output (control) vector |
| $N_{\text{data}}$ | Number of samples in the considered dataset |
| *Controller formulation* | |
| $i$, $k$ | Horizon, sample, or reference index (as specified locally) and combustion-cycle index |
| $u(k)$, $y(k)$ | Engine control-input and output vectors |
| $\Delta u_i$ | Control-input change at horizon index $i$ |
| $N$ | MPC prediction-horizon length |
| $J$ | MPC cost function |
| $Q$, $R$ | Output and input-rate weighting matrices |
| $q_i$, $r_i$ | Diagonal elements of $Q$ and $R$ |
| $r_{\tilde{u}}$ | Input-change weighting in the cost function |
| $s_i$, $q_s$ | Soft-constraint slack vector and its $\ell_1$ penalty weight |
| *Engine quantities* | |
| $p_{\text{mi}}$, $c_{\text{NOx}}$, $c_{\text{PM}}$, $dp_{\text{max}}$ | Engine outputs: indicated mean effective pressure, nitrogen-oxide concentration, particle-mass concentration, and maximum pressure rise rate |
| $p_{\text{mi,ref}}$ | Indicated-mean-effective-pressure reference |
| $t_{\text{Main}}$, $t_{\text{P2M}}$, $\alpha_{\text{Main}}$, $t_{\text{H2}}$ | Control inputs: main-diesel duration, pre-to-main interval, main-diesel start angle, and hydrogen duration |
| $t_{\text{Pre}}$ | Diesel pre-injection duration |
| $p_{\text{Rail}}$, $\alpha_{\text{H2}}$ | Diesel rail pressure and hydrogen start of injection |

## Appendix A. Engine Parameters

Table A.3: Cummins QSB 4.5 Tier 3 parameters, as tested

| **Parameter** | **Value** |
|---|---|
| Total Displacement | 4.460 L |
| Number of Cylinders | 4 |
| Bore | 107 mm |
| Stroke | 124 mm |
| Connecting Rod Length | 192 mm |
| Compression ratio | 17.2:1 |
| Piston Protrusion | 0.43 mm |
| Headgasket Thickness | 1.6 mm |
| Valve train | Pushrod |
| No. of valves (In/Ex) | 2/2 |
| Max. valve lift (In/Ex) | 8 mm/8 mm |
| Valve diameter (In/Ex) | 33 mm/33 mm |
| Combustion Chamber | Bowl in Piston |
| Injection Type | Direct |
| Injection Pressure | 300–1500 bar |
| Spray Angle | 124° |
| Injector Holes | 8 |
| Injector Actuation | Solenoid |

## Appendix B. DNN Training Settings and Diagnostics

Table B.4: Training hyperparameter settings

| **Parameter** | **Value** |
|---|---|
| Max. epochs | 5000 |
| Optimizer | Adam |
| Training metric | RMSE |
| Mini-batch size | 512 |
| Initial learning rate | 0.0005 |
| Learning-rate schedule | Piecewise drop by 25% every 250 epochs |
| L2 regularization | 0.1 |
| Validation frequency | 10 |

## Appendix C. Additional Validation Traces

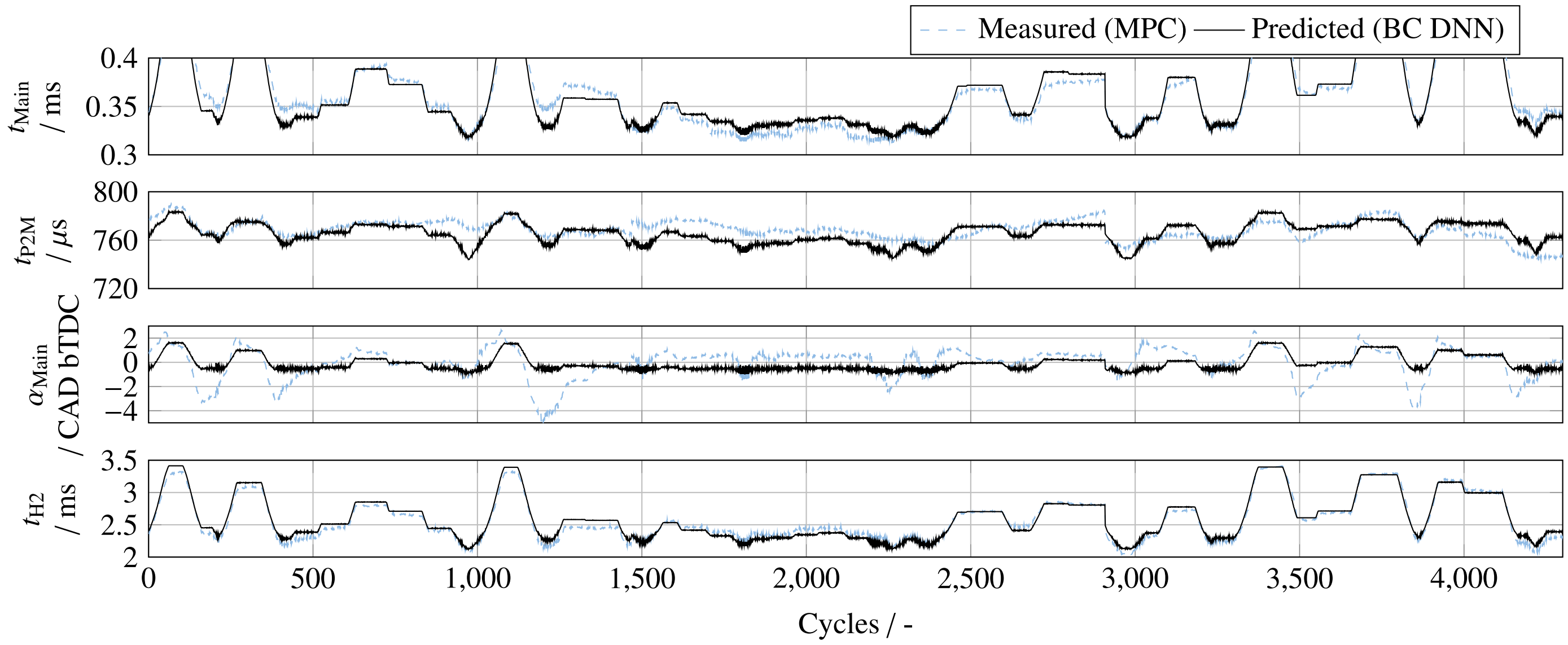

Figure C.10: Predicted DNN outputs on the unseen test dataset over time domain.

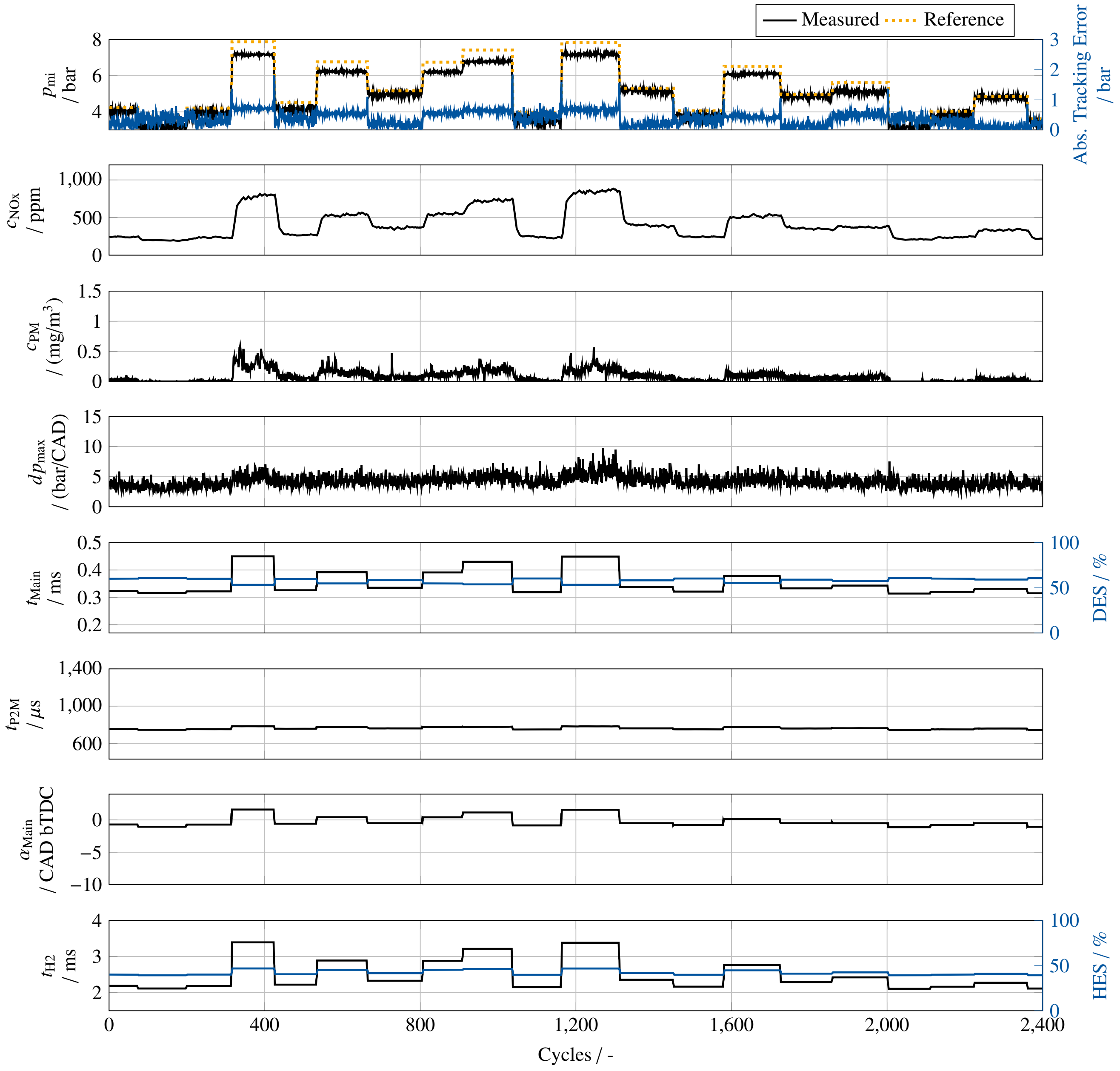

Figure C.11: Process outputs and controls for the BC controller without feedback (Variant B) on an unseen dynamic reference trajectory.